\documentclass[pdflatex,sn-mathphys-num]{sn-jnl}

\usepackage{graphicx}%
\usepackage{multirow}%
\usepackage{slashed}
\usepackage{bm}
\usepackage{cancel}
\usepackage{float}
\usepackage{mathtools}
\usepackage{amsmath,amsfonts,amssymb,amscd,amsxtra,amsthm}
\usepackage{amsbsy}
\usepackage{amstext}

\usepackage[utf8]{inputenc} 
\usepackage{mathrsfs}%
\usepackage[title]{appendix}%
\usepackage{xcolor}%
\usepackage[normalem]{ulem}%
\usepackage{textcomp}%
\usepackage{manyfoot}%
\usepackage{booktabs}%
\usepackage{algorithm}%
\usepackage{algorithmicx}%
\usepackage{algpseudocode}%
\usepackage{listings}%

\theoremstyle{thmstyleone}%
\theoremstyle{thmstyletwo}%

\theoremstyle{thmstylethree}%

\begin{document} 
 
\title[Article Title]{Flavor decomposition of the gravitational form factors and mechanical structure of the proton} 
 
\author*[1]{\fnm{Hyun-Chul} \sur{Kim}}\email{hchkim@inha.ac.kr}

\author*[1]{\fnm{June-Young} \sur{Kim}} 

\author*[2]{\fnm{Ho-Yeon} \sur{Won}} 
 
\affil*[1]{\orgdiv{Department of Physics and Institute of Quantum 
    Science}, \orgname{Inha University}, \orgaddress{\street{100 
      Inha-Ro}, \city{Incheon},  
    \postcode{22212}, \country{Republic of Korea}}}

\affil*[2]{\orgdiv{CPHT, CNRS, \'Ecole polytechnique},
  \orgname{Institut Polytechnique de Paris}, \city{Palaiseau},  
    \country{France}}
 
\abstract{We review in the present talk a series of recent works on the flavor
decomposition of the gravitational form factors of the proton
and its mechanical properties within the framework of the chiral
quark-soliton model. Starting from the energy-momentum tensor 
operator derived from the QCD instanton vacuum, we carry out the twist
projection of the energy-momentum tensor operator into its twist-2 and
twist-4 components, which enables us to isolate the $\overline{c}$
form factor originating from the twist-4 operator. We present the 
flavor-decomposed mass, spin, pressure, and shear-force distributions
of the proton, together with the corresponding form factors. While the
up quark dominates both the mass and the spin of the proton, the
strange quark is found to contribute sizably to the $D$-term form
factor. We also discuss the mechanical stability of the proton
governed by the pressure and shear-force distributions.}  
 
 
\maketitle 
 
\section{Introduction}\label{sec1} 
The gravitational form factors (GFFs) of the nucleon carry information 
on its internal structure that is as essential as that provided by the 
electromagnetic form factors (EMFFs), since they encode the mass, 
spin, and mechanical properties of the nucleon. The GFFs are probed by 
the graviton, as was already suggested in the 
1960s~\cite{Kobzarev:1962wt, Pagels:1966zza}. Since the graviton 
interacts with the nucleon far too weakly to be measured, however, the 
GFFs had long remained objects of purely theoretical interest, until 
the advent of generalized parton distributions (GPDs) in the 1990s 
opened experimental access to them through hard exclusive processes 
such as deeply virtual Compton scattering (DVCS)~\cite{Muller:1994ses, 
  Ji:1996ek, Radyushkin:1996nd}. The EMFFs and GFFs are then obtained 
respectively as the first and second Mellin moments of the GPDs with
respect to the parton momentum fraction $x$. Recently, Burkert et
al. reported the first extraction of the $D$-term form factor from DVCS 
data~\cite{Burkert:2018bqq, Kumericki:2019ddg, Burkert:2021ith}. 
Moreover, the Electron-Ion Collider (EIC) is currently being 
constructed with three key scientific goals: to understand the origin 
of the nucleon mass, the origin of the nucleon spin, and the emergent 
properties of dense systems of gluons~\cite{Accardi:2012qut, 
  AbdulKhalek:2021gbh}. 
 
Since the nucleon GFFs are defined through the 
parametrization of the nucleon matrix element of the energy-momentum 
tensor (EMT) operator~\cite{Kobzarev:1962wt, Pagels:1966zza, 
Polyakov:2018zvc}, we need to construct this operator from 
the QCD action. Varying the QCD action coupled to a weak classical 
torsionless gravitational field with respect to the metric 
$g_{\mu\nu}$, we can derive the Belinfante-Rosenfeld-type QCD EMT 
operator:  
\begin{align} 
T^{\mu\nu} = \sum_q T_q^{\mu\nu} + T_g^{\mu\nu}    , 
\end{align} 
where $T_q^{\mu\nu}$ and $T_g^{\mu\nu}$ denote respectively the quark 
and gluon parts 
\begin{align} 
    T^{\mu \nu}_{q}   
& = \frac{i}{4}  \bar{\psi}_{q}  \left(  
    \gamma^{ \{\mu} \overleftrightarrow{\mathcal{D}}^{\nu \}} 
    \right)    \psi_{q},  
    \quad     
    T^{\mu \nu}_{g}   
  = -G^{ \mu \rho,a} G^{\nu,a  }_{\ \rho}  
  + \frac{1}{4} g^{\mu \nu} G^{ \lambda \rho, a} G^{a}_{\lambda \rho}  
  \label{eq:1}   
\end{align} 
Here, $\overleftrightarrow{\mathcal{D}}^{\mu} = 
\overleftrightarrow{\partial}^{\mu} - 2 i g A^{\mu}$ represents the 
covariant derivative with $\overleftrightarrow{\partial}^{\mu} =  
\overrightarrow{\partial}^{\mu}- \overleftarrow{\partial}^{\mu}$, and 
$a^{ \{ \mu }b^{ \nu \} } = a^{\mu} b^{\nu} + a^{\nu} 
b^{\mu}$. $G^{\mu \nu, a}$ stands for the gluon field strength with the 
color index $a$. Note that while the total EMT operator is conserved, its quark and 
gluon parts separately are not:
\begin{align}
  \label{eq:2}
\partial _\mu T^{\mu\nu} = 0,\;\;\; \partial _\mu T_q^{\mu\nu} \neq 0,
  \;\;\; \partial _\mu T_g^{\mu\nu} \neq 0 .
\end{align}
This implies that the quark and gluon contributions are individually 
scale-dependent. When the two contributions are summed, however, the  
scale dependence cancels, so that the total EMT operator becomes 
scale-independent.

The matrix element of the EMT current can be parametrized 
by four independent Lorentz-invariant form factors:
\begin{align}
&   \langle N(p',J'_{3}) |    T_{\mu\nu}^{a}  ( 0 )
                |N (p,J_{3})\rangle 
  = \bar{u}(p',J'_{3})
    \Bigg[
    A^{a}  ( t ) \frac{  P_{\mu} P_{\nu} }{  M_N } 
  + J^{a}  ( t ) \frac{  i  P_{ \{ \mu} \sigma_{\nu \} \rho}
  \Delta^{\rho} } { 2 M_N }  \cr 
&   \hspace{4.2cm}
  + D^{a}  ( t )  \frac{ \Delta_{\mu} \Delta_{\nu} -  g_{\mu\nu}
  \Delta^{2}}{4M_N}      
  + \overline{c}^{a} ( t ) M_N g_{\mu\nu}  
    \Bigg] 
    u(p, J_{3}),
\label{eq:4}
\end{align}
where $A^{a}$, $J^{a}$, $D^{a}$, and $\overline{c}^{a}$ are referred to as
the energy, angular momentum, $D$-term, and $\overline{c}$ form
factors of the nucleon, respectively. Note that the $\overline{c}^{a}$
form factor vanishes as a consequence of the conservation of the total
EMT current. When the quark and gluon parts are treated separately,  
however, the corresponding $\overline{c}^{a}$ form factors are 
non-vanishing. The relativistic normalization of the one-particle 
nucleon state is given by
$\langle N' (p', J'_{3})| N (p, J_{3}) \rangle = 2p^{0} (2\pi)^{3}
\delta_{J'_{3} J_{3}} \delta^{(3)}(\bm{p}'-\bm{p})$, 
where $J_{3}$ and $J'_{3}$ denote the spin polarizations of the 
initial and final states, respectively. $M_{N}$ stands for the 
nucleon mass, while $p$ and $p'$ represent the initial and final 
momenta. The average momentum and the momentum transfer are defined  
as $P=(p'+p)/2$ and $\Delta=p'-p$, respectively, with 
$\Delta^{2}=t$. 

Let us denote the GFFs generically by $F^{\chi}$, where the flavor 
index $\chi$ takes the values $\chi=0,3,8$. The flavor singlet, 
triplet, and octet GFFs are then expressed in terms of the 
flavor-decomposed GFFs as
\begin{align}
  F^{\chi = 0} =F^{u}+F^{d}+F^{s}, \quad
  F^{\chi = 3} = F^{u}- F^{d}, \quad
  F^{\chi = 8} = \frac{1}{\sqrt{3}}\left(F^{u}+F^{d} -2F^{s}\right).  
\label{eq:5}
\end{align}
For the vector currents, the flavor singlet, triplet, and octet
components can be readily constructed with the aid of the Gell-Mann
matrices $\lambda^{\chi}$. For the flavor EMT currents, however, the
construction is considerably more involved.

In the chiral-quark-soliton model ($\chi$QSM), which is the
theoretical framework employed to analyze the GFFs of the nucleon in
the current talk, the \emph{effective} EMT operator~\cite{Won:2023zmf}
can be written as
\begin{align}
  \label{eq:6}
T^{\mu \nu}_{q} (x) + T^{\mu \nu}_{g} (x)~[\mathrm{QCD}]
  \xrightarrow{\mathrm{eff}} T^{\mu \nu,\chi=0}
  (x) = \frac{i}{4}  \bar{\psi}  \left(   
    \gamma^{ \{\mu} \overleftrightarrow{\partial}^{\nu \}} 
    \right)    \psi,
\end{align}
where the gluon degrees of freedom have been integrated out from the
instanton vacuum. The N\"other theorem does not provide
any separate information on the gluon contribution, since all
nonperturbative gluon effects are already encoded in the effective
EMT operator. One might naively construct the 
flavor currents by inserting the flavor SU(3) Gell-Mann matrices:
\begin{align}
    T^{\mu\nu}_{\chi}(x) 
  = \frac{i}{4}   \bar{\psi} (x)
    \gamma^{ \{\mu} \overleftrightarrow{\partial}^{\nu\} }
    \lambda_{\chi} 
    \psi (x).
\label{eq:7}
\end{align}
Note that the trace-anomaly contribution (twist-4) is excluded by
projecting onto the twist-2 operator.
Once the twist-4 part is removed, the only remaining gluonic
contribution is that of the twist-2 gluon operator. In the dilute
instanton vacuum, the nucleon matrix elements of the twist-2 quark
operators are of order unity, whereas those of the twist-2 gluon
operator arise solely from the instanton fields and are parametrically
suppressed by the small packing fraction
$(\bar{\rho}/\bar{R})^{4}\sim 10^{-2}$~\cite{Diakonov:1996sr,
Diakonov:1997vc}. The identification of Eqs.~\eqref{eq:6}
and~\eqref{eq:7} with the QCD twist-2 quark operators holds at the low
normalization point $\mu_{0}\sim\bar{\rho}^{-1}\approx 600$~MeV and to
leading order in the packing fraction, at which all nonperturbative
gluonic effects are resummed into the chiral background field.
Although the Dirac operator of the $\chi$QSM contains an explicit
coupling to this background field, the gluon-field part of the QCD
covariant derivative contributes to the matrix elements of the
leading-twist quark operators only at subleading order, so that
ordinary derivatives suffice for the twist-2 operators considered
here. For higher-twist operators, in contrast, the gluon-field parts
of the covariant derivatives contribute at leading order and cannot be
neglected~\cite{Balla:1997hf, Polyakov:2018exb}.

In fact, we have already performed the flavor decomposition of
the GFFs~\cite{Won:2023cyd, Won:2023ial} by using Eq.~\eqref{eq:7},
and the same prescription was also employed in
Refs.~\cite{Wakamatsu:2005vk, Wakamatsu:2006dy, Wakamatsu:2007uc}.
Strictly speaking, however, the flavor-triplet and -octet EMT
operators cannot be constructed naively by inserting the Gell-Mann
matrices, as was done in the electromagnetic case. While the
flavor-singlet EMT current~\eqref{eq:6} can be regarded as a
N\"other current, a flavor-dependent EMT current is in general not
conserved and hence is not a N\"other current. Thus, there is no
theoretically rigorous way to derive the flavor-dependent EMT currents
within an effective theory. Nevertheless, it is of great importance to
construct the flavor-nonsinglet EMT currents such as
$T_{\chi}^{\mu\nu}$. In the present talk, we will focus on the
flavor decomposition of the GFFs within the framework of the
$\chi$QSM, keeping in mind that the gluonic contribution is no longer
suppressed beyond the leading twist (twist~2). 
For the flavor-nonsinglet sector, the twist-4 operators vanish at the
classical level by the QCD equations of motion, and no
quantum-fluctuation contribution arises to make them
nonzero. Consequently, the twist-2 projection effectively represents
the full EMT operator in the flavor-nonsinglet sector. 

\section{Twist projection of the effective EMT operators} 
Extracting the $\overline{c}$ form factor is a difficult task in most
dynamical models and lattice QCD due to the higher-twist (twist-4)
corrections. To isolate this twist-4 term, we need to 
decompose the symmetrized EMT current in terms of the twist-2 and
twist-4 components: 
\begin{align}
  T^{\mu \nu} =  \bar{T}^{\mu \nu} + \hat{T}^{\mu \nu},
  \label{eq:8}
\end{align}
 where $\bar{T}^{\mu \nu}$ and $\hat{T}^{\mu \nu}$ are called
respectively the twist-2 and twist-4 EMT operators
\begin{align}
    \bar{T}^{\mu \nu}  
 = T^{\mu \nu}  - \frac{1}{4} g^{\mu \nu}  T^{\alpha}_{\alpha},
  \quad     \hat{T}^{\mu \nu}   
 = \frac{1}{4} g^{\mu \nu}  T^{\alpha}_{\alpha}.
\end{align}
The nucleon matrix elements of the twist-2 and twist-4 EMT operators
are given respectively as 
\begin{align}
&   \langle N(p',J'_{3}) |    \bar{T}_{\mu\nu}  ( 0 )
                |N(p,J_{3})\rangle               \cr
  &= \bar{u}(p',J'_{3}) \Bigg[
    A  ( t ) \frac{  P_{\mu} P_{\nu} }{  M_N } 
  + J  ( t ) \frac{  i  P_{ \{ \mu} \sigma_{\nu \} \rho}
  \Delta^{\rho} } { 2 M_N }  
  + D  ( t )  \frac{ \Delta_{\mu} \Delta_{\nu} -  t g^{\mu \nu} }{4M_N}      
 \cr
&\hspace{0.4cm} - g_{\mu\nu}  
    \left\{ \frac{t}{8 M_{N}} J(t) 
  - \frac{3t}{16 M_{N}} D(t) 
  + \frac{M_{N}}{4}  
    \left(1- \frac{t}{4M_{N}^{2}}\right)
    A (t) \right\}
       \Bigg]     u(p,J_{3}),
\label{eq:10}       
\end{align}
and
\begin{align}
&   \langle {N(p',J'_{3})}|    \hat{T}_{\mu\nu}  ( 0 )
                |{N(p,J_{3})}\rangle \cr 
& = \bar{u}(p',J'_{3})
    \Bigg[ g_{\mu\nu}  \bigg{\{} 
    M_{N} \overline{c}(t)
  + \frac{t}{8 M_{N}} J(t) \cr
&\hspace{2.5cm}- \frac{3t}{16 M_{N}} D(t) 
  + \frac{M_{N}}{4}  
    \left(1- \frac{t}{4M_{N}^{2}}\right)
    A (t) \bigg{\}}
    \Bigg]   u(p,J_{3}).
\label{eq:11}                                            
\end{align}
Examining Eq.~\eqref{eq:11}, we find that the $\overline{c}$ form factors
arise from the twist-4 EMT operator. 
The sum of the above twist-2 and twist-4 parts must be the same as the
parametrization given in Eq.~\eqref{eq:4}:  
\begin{align}
& \langle{N(p',J'_{3})}|    {T}_{\mu\nu}  ( 0 )  |
                {N(p,J_{3})}\rangle =\cr 
&   \hspace{2cm}
    \langle{N(p',J'_{3})}|    \bar{T}_{\mu\nu}  ( 0 )  
    |{N(p,J_{3})}\rangle +\langle{N(p',J'_{3})}|
        \hat{T}_{\mu\nu}  ( 0 )   
    |{N(p,J_{3})}\rangle.
\end{align}

\section{Results and discussion}
\begin{figure}[htp] 
\centering 
\includegraphics[width=0.49\textwidth]{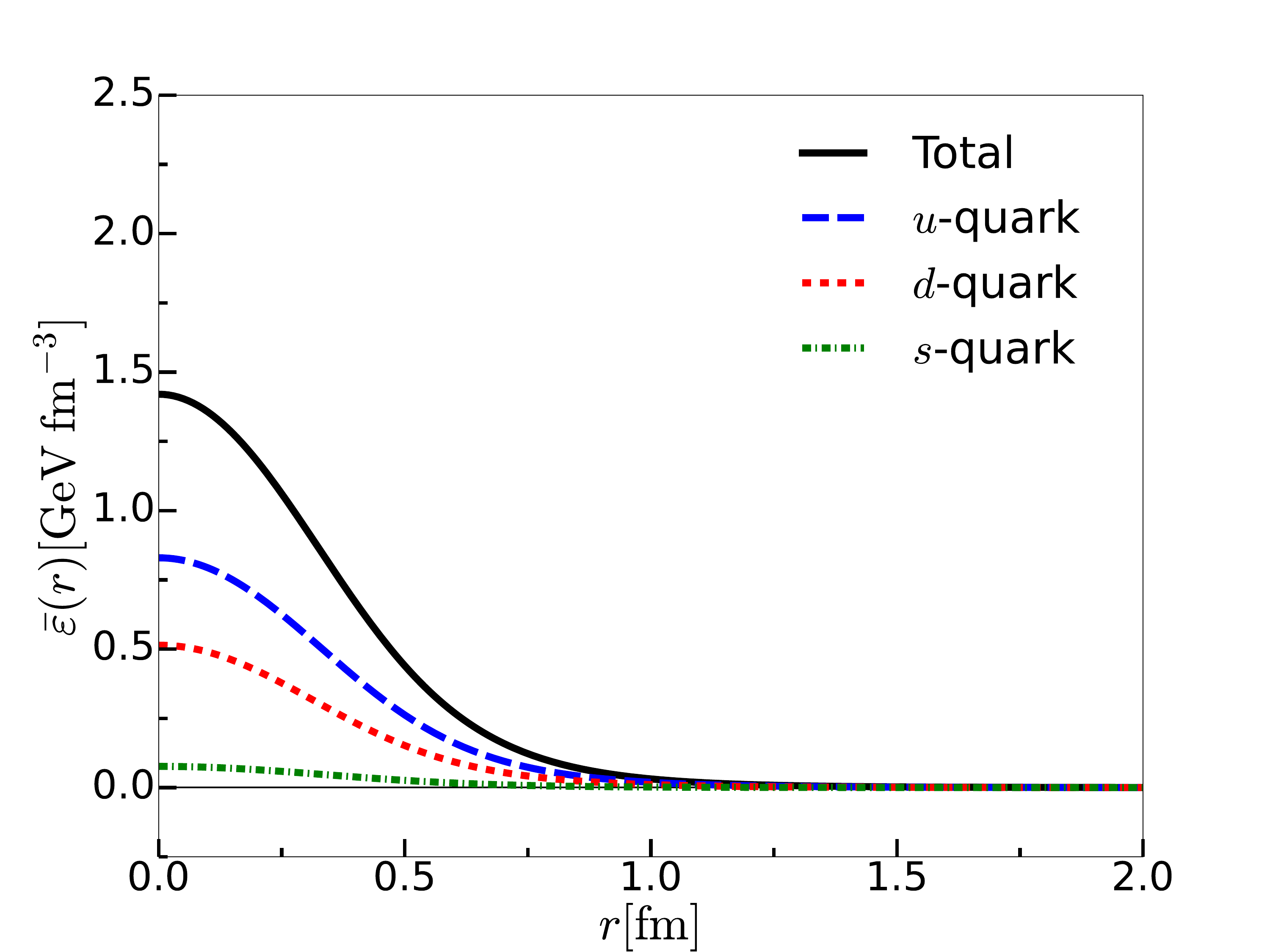}
\includegraphics[width=0.49\textwidth]{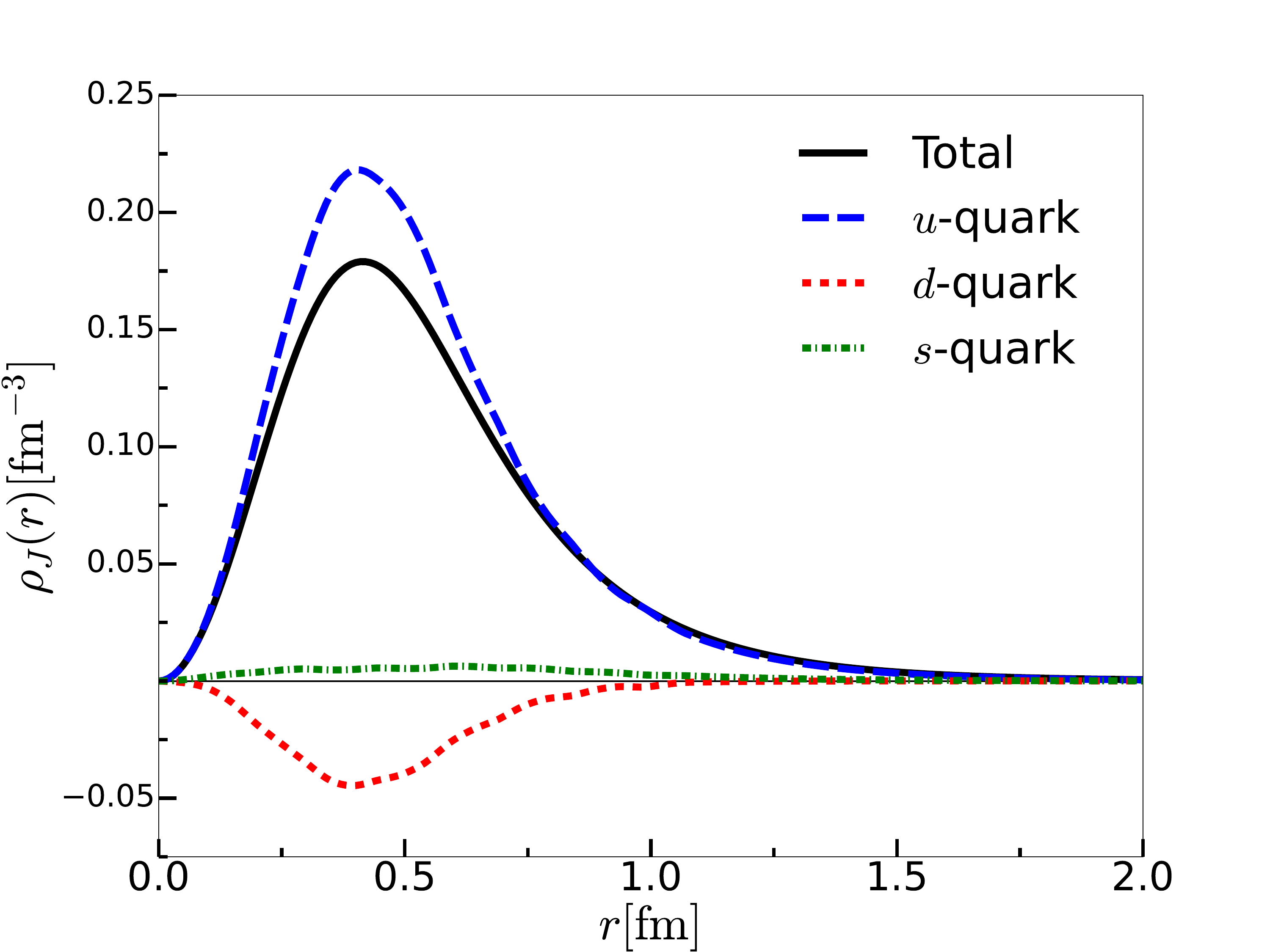}
\includegraphics[width=0.49\textwidth]{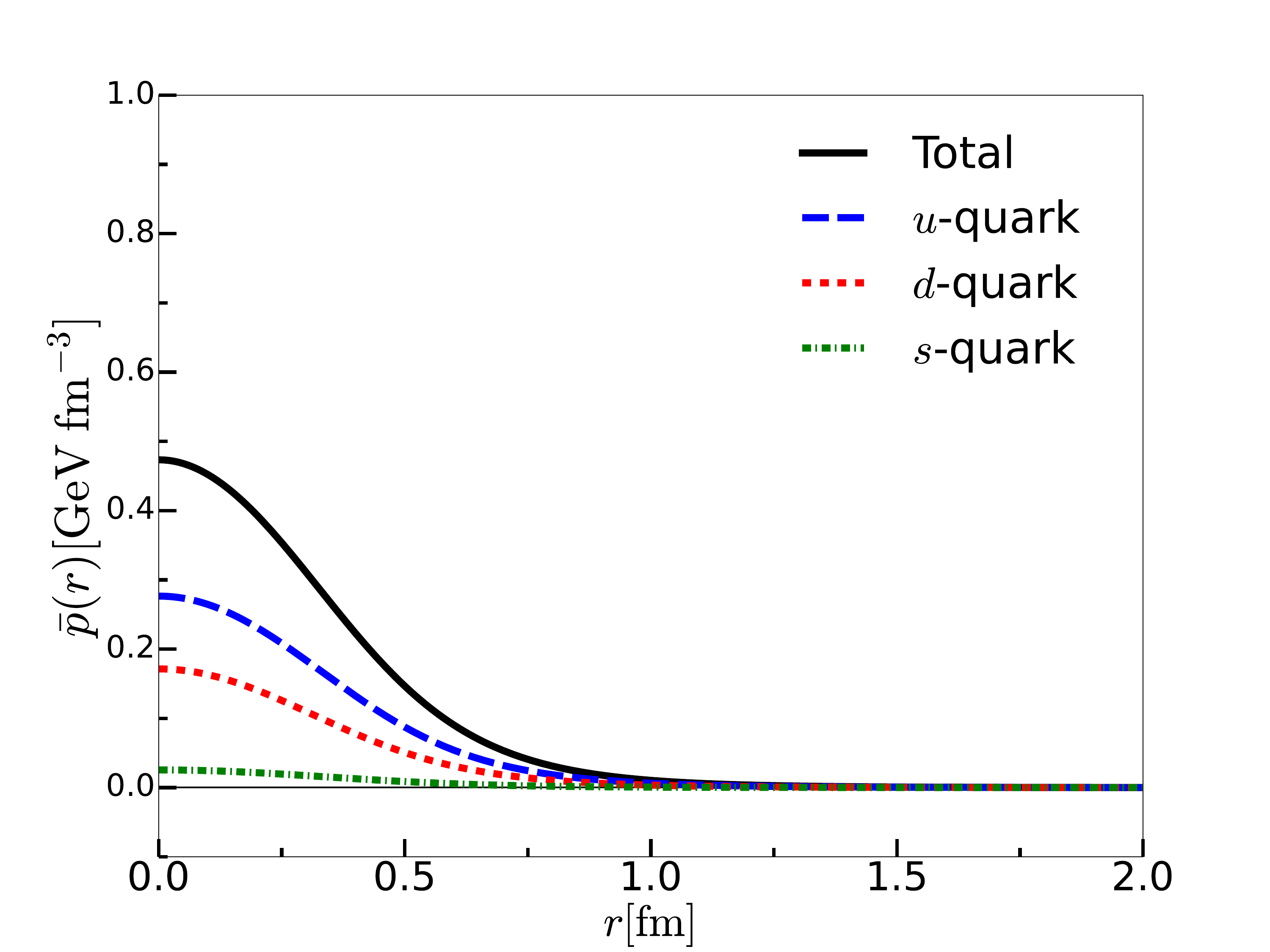}
\includegraphics[width=0.49\textwidth]{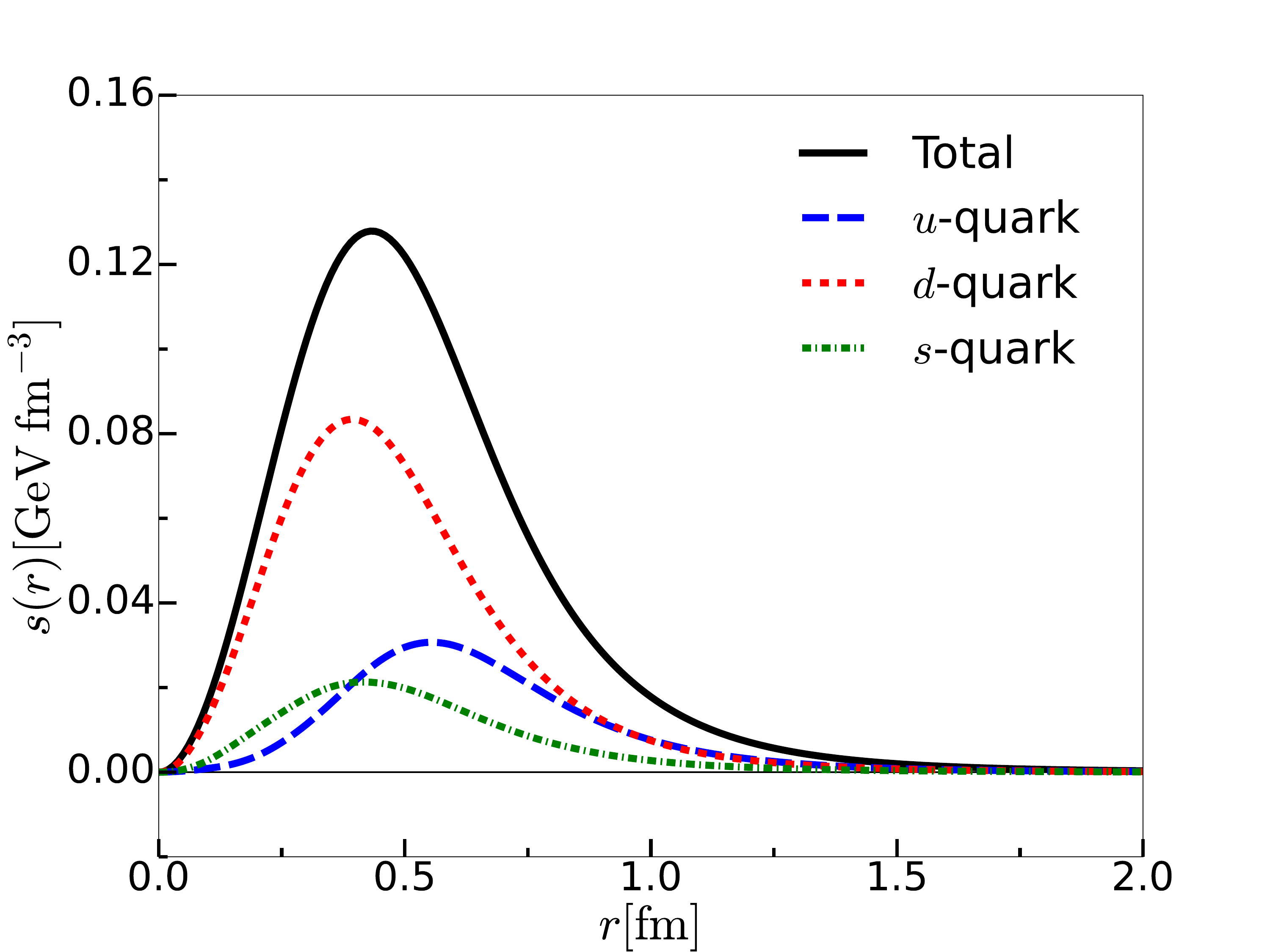} 
\caption{The twist-2 flavor-decomposed mass distribution, the
  flavor-decomposed spin distribution, the twist-2 flavor-decomposed
  pressure distribution, and the flavor-decomposed shear-force
  distribution, shown from the upper-left to the lower-right panel,
  respectively.
}
\label{fig:1}  
\end{figure}
\begin{figure}[htp] 
\centering 
\includegraphics[width=0.49\textwidth]{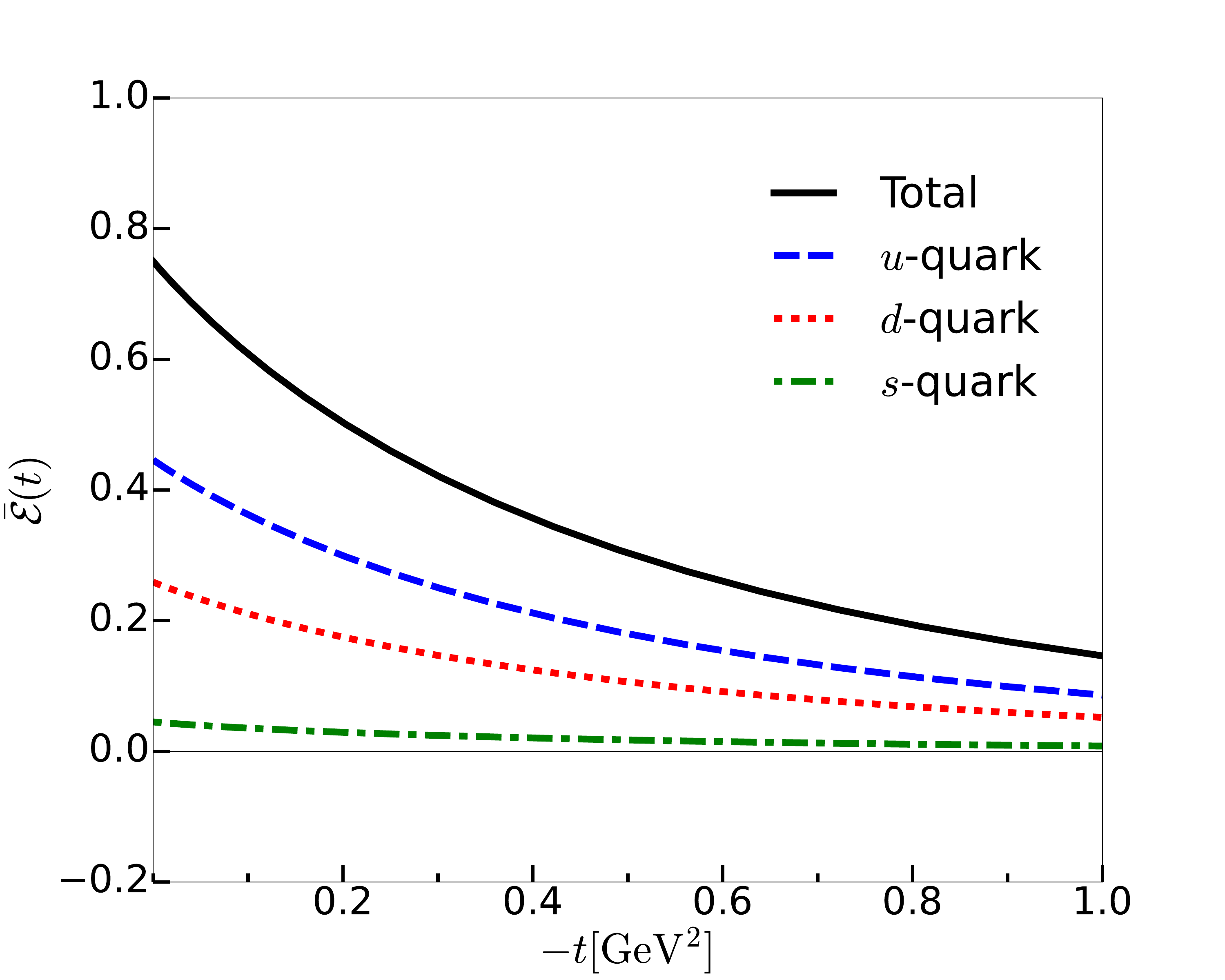}
\includegraphics[width=0.49\textwidth]{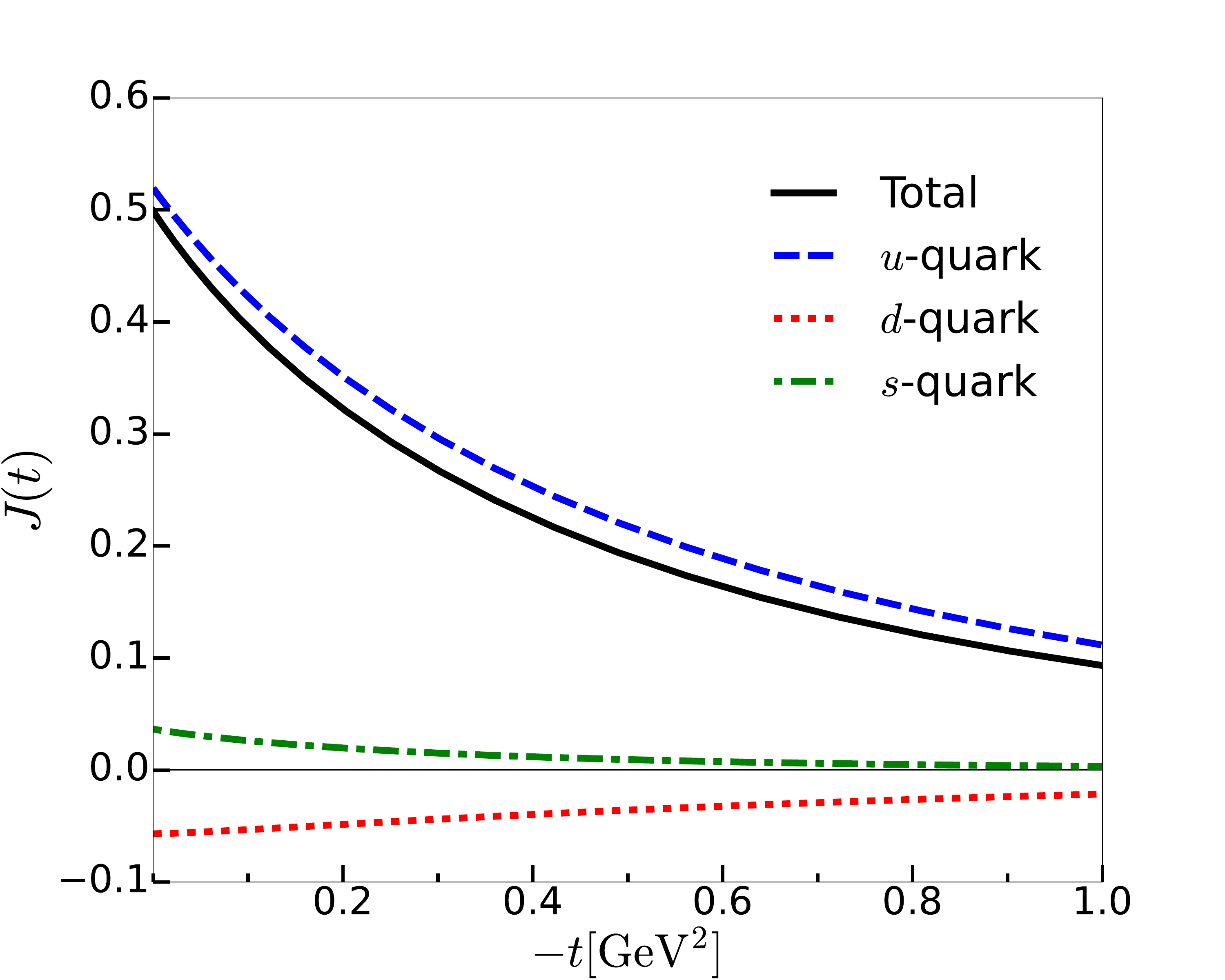}
\includegraphics[width=0.55\textwidth]{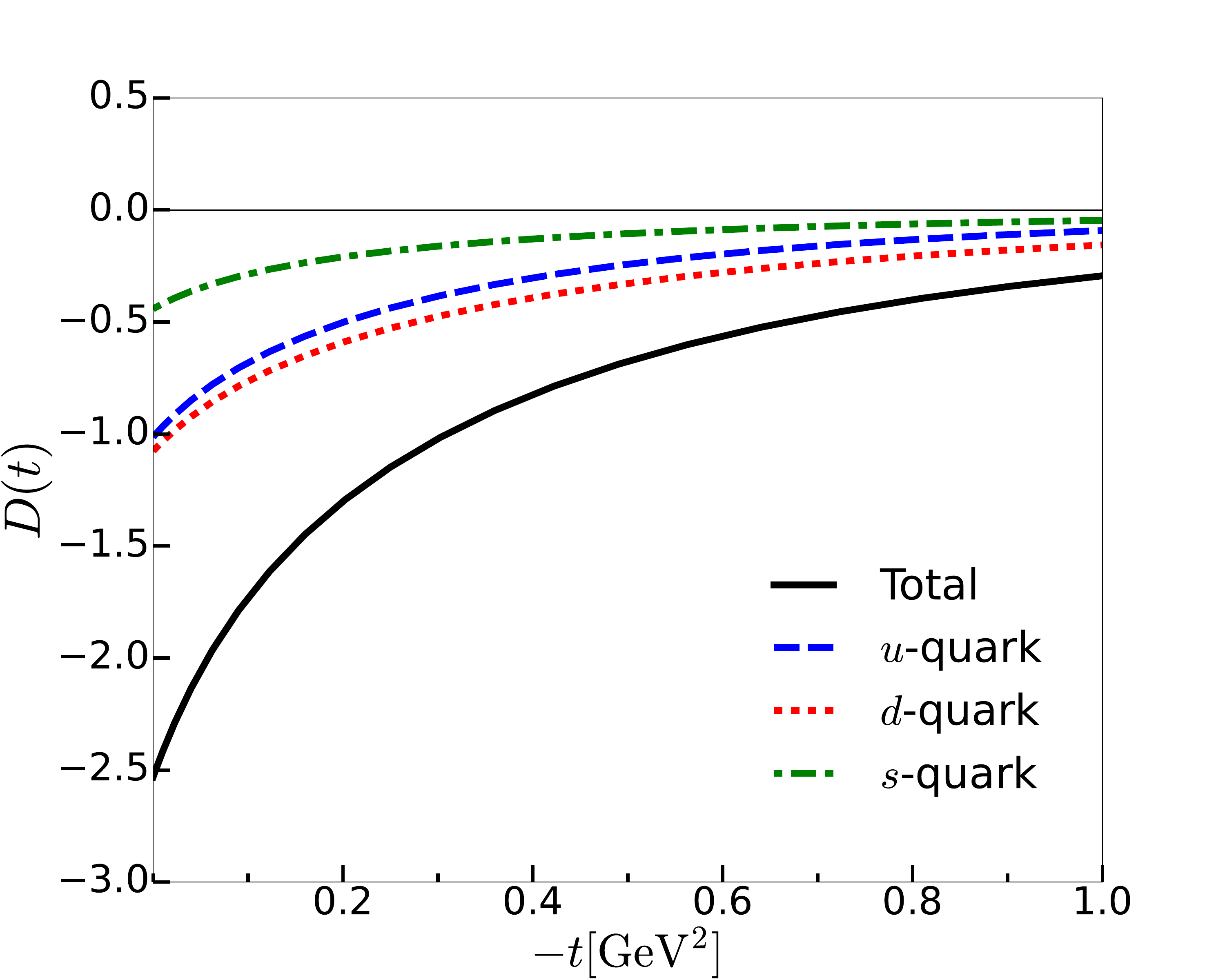}
\caption{The twist-2 flavor-decomposed energy form factors, the
  flavor-decomposed spin form factors, and the flavor-decomposed
  $D$-term form factors, shown from the upper-left to the lower panel, 
  respectively.}
\label{fig:2} 
\end{figure}

As shown in Eq.~\eqref{eq:5}, we can perform the flavor decomposition
of the EMT densities and the corresponding GFFs. In the upper-left
panel of Fig.~\ref{fig:1}, we display the twist-2 flavor-decomposed
energy distribution for the nucleon. Integrating the energy distribution
yields the nucleon mass:
\begin{subequations}
\begin{align}
   & \int d^{3}r\, \varepsilon^f (r)  
  = M_{N} [A^f(0) + \overline{c}^f(0) ], \label{eq:13a} \\
   & \int d^{3}r\, \bar{\varepsilon}^f (r)  
= \frac{3}{4} M_{N} A^f (0),
\label{eq:13b} \\
   & \int d^{3}r\, \hat{\varepsilon}^f (r)  
     = \frac{1}{4} M_{N} [A^f (0) + 4  \overline{c}^f(0)  ],
\label{eq:13c}
\end{align}
\end{subequations}
where the superscript $f$ represents the flavor component,
i.e. $f=u,\,d,\,s$. Note that the twist-2 energy distribution carries
three fourths of the nucleon mass, as given in Eq.~\eqref{eq:13b}. The
$\overline{c}^f$ term originates from the twist-4 contribution, as
indicated by Eq.~\eqref{eq:13c}. Although the total $\overline{c}$ vanishes
owing to the conservation of the effective EMT current, each
flavor-decomposed component survives. As seen in the upper-left panel
of Fig.~\ref{fig:1}, the up-quark contribution dominates the proton
mass, as expected, whereas the strange-quark contribution is tiny. In
the upper-left panel of Fig.~\ref{fig:2}, we observe that the total 
twist-2 mass form factor is normalized to $3/4$ of the nucleon mass,
as already noted in Eq.~\eqref{eq:13b}.
 
The angular momentum (AM) distribution~\cite{Polyakov:2002yz} can be
defined as 
\begin{align}
  \rho_{J}(r) = \frac{1}{3}\sum_{J'_3
  J_3}\varepsilon^{ijk} (J_i)_{J_3 J'_3} x_j T_{0k}(r,  J_3, J_3')  
\label{eq:14}
\end{align}
where 
\begin{align}
T_{0k}(r, J_3, J_3') = \int \frac{d^3 \Delta}{(2\pi)^32E}
e^{i\bm{r} \cdot \bm{\Delta}}   
  \langle{N(p',J'_{3})}|    {T}_{0k}  ( 0 )  | {N(p,J_{3})}\rangle.
    \label{eq:15}
\end{align}
We consider only the monopole distribution $\rho_J$.
Since $\rho_{J}$ arises from the off-diagonal $0k$ components of the
EMT operator, the twist-4 part does not contribute to the spin
distribution. The spatial integral of the AM distribution
$\rho_{J}$ yields the total spin of the nucleon, i.e., $1/2$. The
upper-right panel of Fig.~\ref{fig:1} shows the numerical results for
the flavor-decomposed spin distributions. Again, the up quark
contributes dominantly to the proton spin, whereas the down quark is
oppositely polarized. The strange quark inside the nucleon barely
contributes to its spin. In the upper-right panel of
Fig.~\ref{fig:2}, the corresponding spin form factors $J^f(t)$ are
displayed. Note that $J(0)$ is exactly equal to $1/2$.

The spatial components of the effective EMT operator, $T_{ij}$, are
related to the mechanical properties of the nucleon. They contain both
the pressure and shear-force distributions inside the nucleon. The
spatial component $T_{ij}$ can be expressed in terms of irreducible
tensors, which correspond to the pressure and shear-force
distributions, respectively:
\begin{align}
T_{ij}(r, J_3, J_3') = 
   p (r) \delta^{ij} \delta_{J'_{3}J_{3}}
  + s (r)  \left(  \frac{r^{i}r^{j}}{r^{2}} - \frac{1}{3}
  \delta^{ij}  \right)  \delta_{J'_{3}J_{3}}.
\label{eq:tij}
\end{align}
The pressure and shear-force distributions are related to the pressure
form factor and the $D$-term form factor, respectively:
\begin{align}
&   p (r)
  =  M_{N}     \int     \frac{d^{3}\Delta}{(2\pi)^{3}}
                e^{-i\bm{\Delta} \cdot  \bm{r}}
                \mathcal{P} (t), \cr 
&   s (r)
  = - \frac{1}{4M_{N}}  r \frac{d}{dr}  \frac{1}{r} \frac{d}{dr}
                                               \int \frac{d^{3}\Delta}{(2\pi)^{3}}
    e^{-i\bm{\Delta}\cdot \bm{r}} D (t). 
\label{eq:psdef}
\end{align}
The pressure form factor receives contributions from both the twist-2
and twist-4 projected EMT operators, which can be written respectively
as
\begin{align}
        \bar{\mathcal{P}} (t) & = \frac{1}{4}
    \left[ A(t) - \frac{t}{4M_{N}^{2} } \left( A(t) - 2 J(t) 
  + \frac{1}{3} D(t)\right)     \right], \cr
    \hat{\mathcal{P}} (t) & = -\frac{1}{4} 
    \left[ A(t)  + 4\overline{c}(t) - \frac{t}{4M_{N}^{2} } 
    \left( A(t) - 2 J (t) + 3 D (t) \right) \right].
\end{align}
We observe again that the $\overline{c}$ form factor appears only in
the twist-4 pressure form factor. The total pressure form factor and
distribution are obtained by adding the twist-2 and twist-4
contributions:
\begin{align}
  \mathcal{P} (t) = \bar{\mathcal{P}} (t) +
  \hat{\mathcal{P}} (t), \quad 
  p(r)=\bar{p} (r)  + \hat{p} (r) .
\label{eq:19}
\end{align}
Thus, the total pressure form factor consists of the $\overline{c}$
and $D$-term form factors:
\begin{align}
  \label{eq:20}
  \mathcal{P} (t) & = - \overline{c} (t)
                    + \frac{t}{6M_{N}^{2} }  D (t) . 
\end{align}
Although the total $\overline{c}$ form factor vanishes, it still
contributes to the flavor-decomposed pressure form factors and
distributions. In the forward limit, we obtain the following
relations:
\begin{align}
    \mathcal{P} (0)  =  - \overline{c} (0) 
, \quad
    \bar{\mathcal{P}} (0)  = \frac{1}{4}   A (0)
    , \quad
    \hat{\mathcal{P}} (0)  = -\frac{1}{4}  A (0) -\overline{c} (0).
\end{align}
The twist-projected 3D multipole energy and pressure form factors are
related to each other:
\begin{align}
    \bar{\mathcal{P}} (t)  = \frac{1}{3} \bar{\mathcal{E}} (t), \quad 
    \hat{\mathcal{P}} (t) = - \hat{\mathcal{E}} (t), 
\end{align}
which leads to the same relations for the energy and pressure
distributions:
\begin{align}
\bar{\varepsilon} (r) = 3 \bar{p} (r), \quad
  \hat{\varepsilon} (r) = - \hat{p} (r). 
\label{eq:23}
\end{align}

In the lower-left panel of Fig.~\ref{fig:1}, the results for the
twist-2 flavor-decomposed pressure distributions are shown. They are
all positive over the entire space, which indicates that the pressure
arising from the twist-2 EMT current is repulsive. Only when the
twist-4 contribution is included does the pressure distribution
satisfy the following stability condition:
\begin{align}
  \int^{\infty}_{0} dr \   r^{2} p (r)=0.
  \label{eq:stability}
\end{align}
The results for the shear-force distributions are displayed in the
lower-right panel of Fig.~\ref{fig:1}. Since the shear force is related
to the off-diagonal spatial component of $T_{ij}$, it is free from the
twist-4 contributions. Together with the pressure distribution, the
shear-force distribution provides a local stability
condition~\cite{Perevalova:2016dln}:
\begin{align}
\frac{2}{3} s(r) + p(r) > 0.
\label{eq:mecstab1}
\end{align}
Note that this condition is associated with the normal force field,
whereas the tangential force is expressed as $-\frac{1}{3} s(r) +
p(r)$. As shown in the lower-right panel of Fig.~\ref{fig:1}, the
shear-force distribution satisfies the positivity condition
$s(r)>0$, so that $\frac{2}{3} s(r) + p(r)$ is a monotonically
decreasing function.

In the lower panel of Fig.~\ref{fig:2}, we present the results for the
flavor-decomposed $D$-term form factors. Interestingly, the
strange-quark contribution to the $D$-term form factor is sizable.
Considering the multipole structure of the GFFs, the $D$-term form
factor has a quadrupole tensor structure. As in the case of the
electric quadrupole form factor of the $\Delta$
isobar~\cite{Kim:2019gka}, the sea-quark contributions come into
significant play, as indicated by the sizable strange-quark
contribution.

\begin{figure}[htp] 
\centering 
\includegraphics[width=0.49\textwidth]{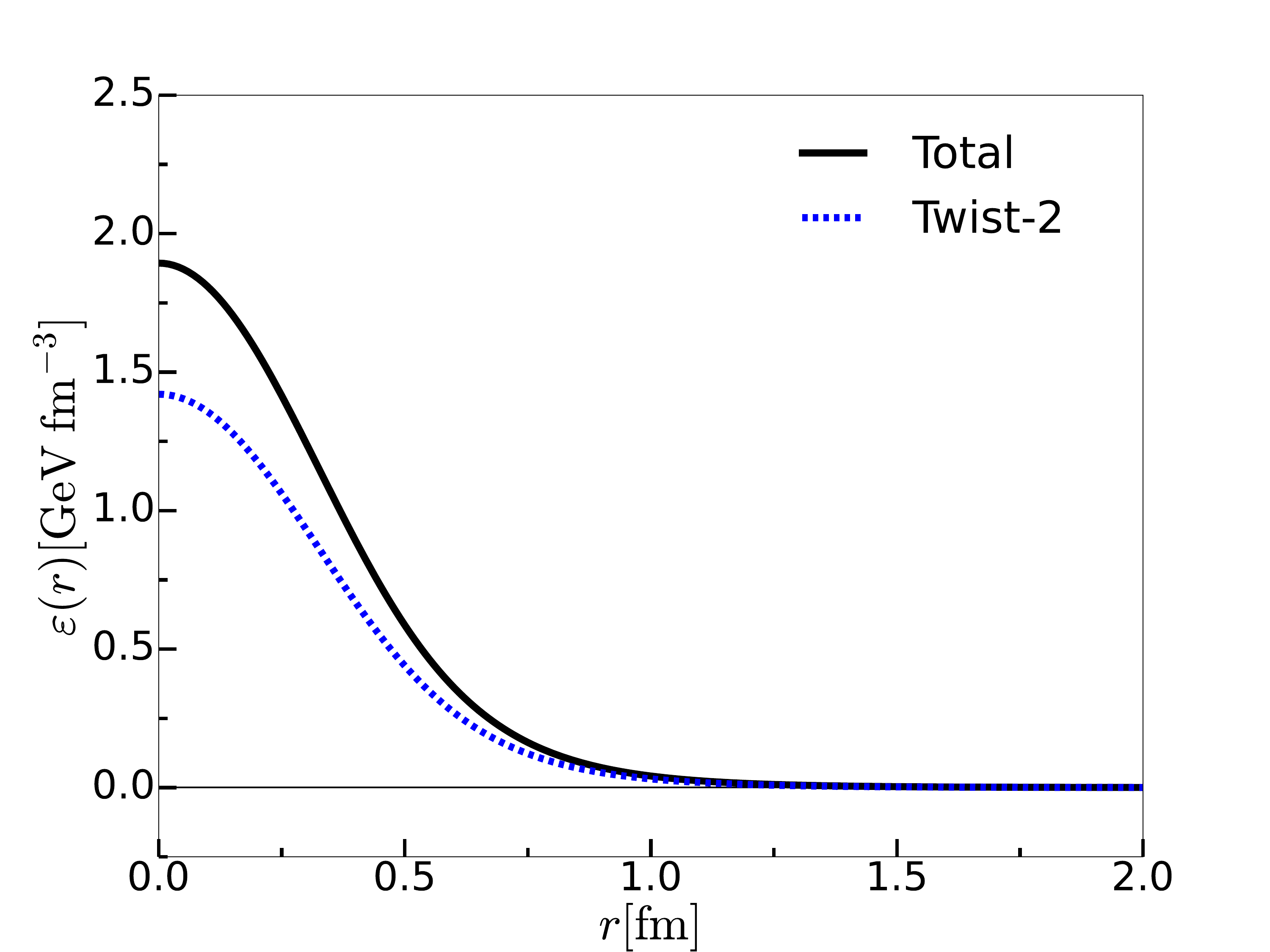}
\includegraphics[width=0.49\textwidth]{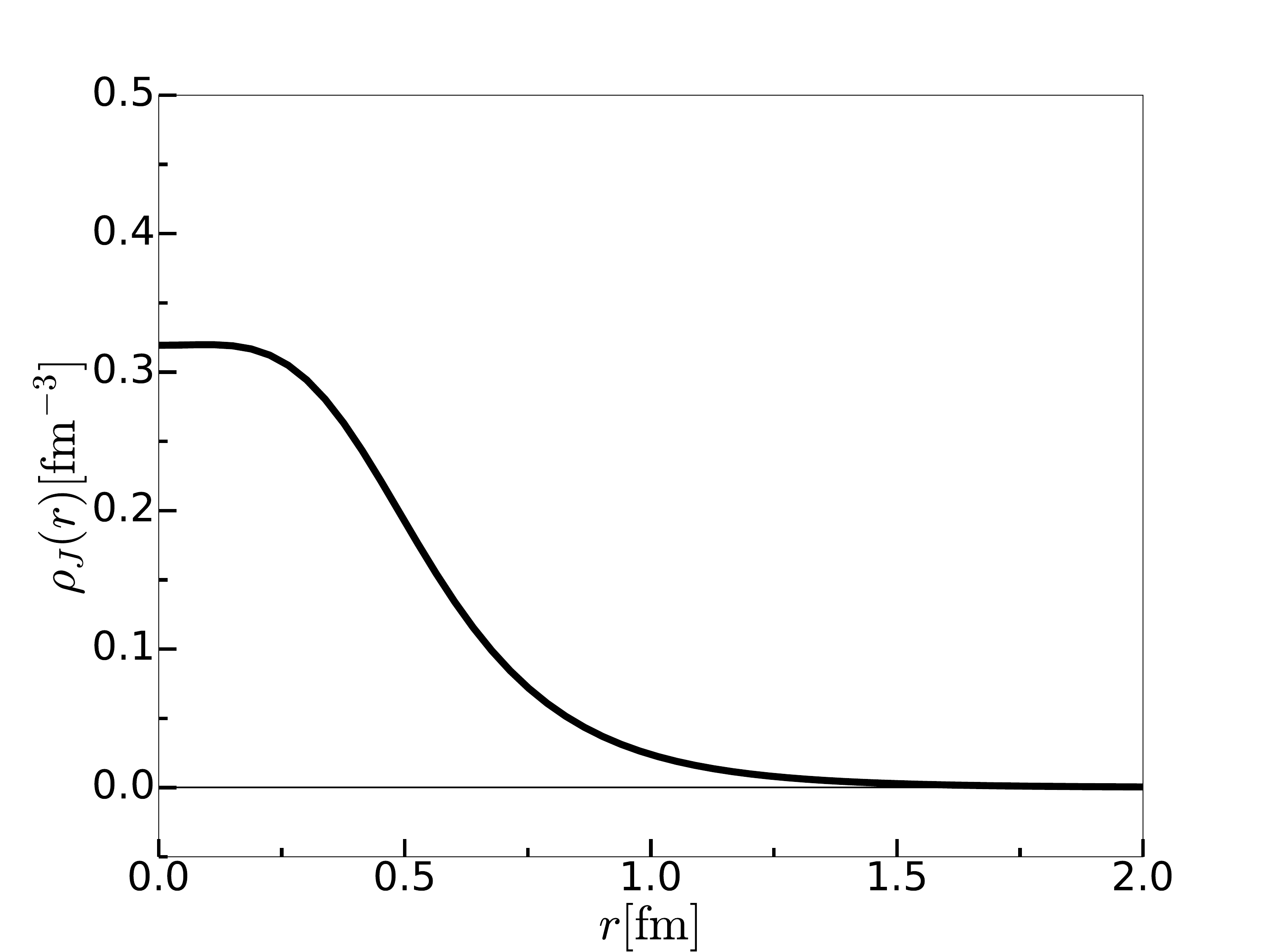}
\includegraphics[width=0.49\textwidth]{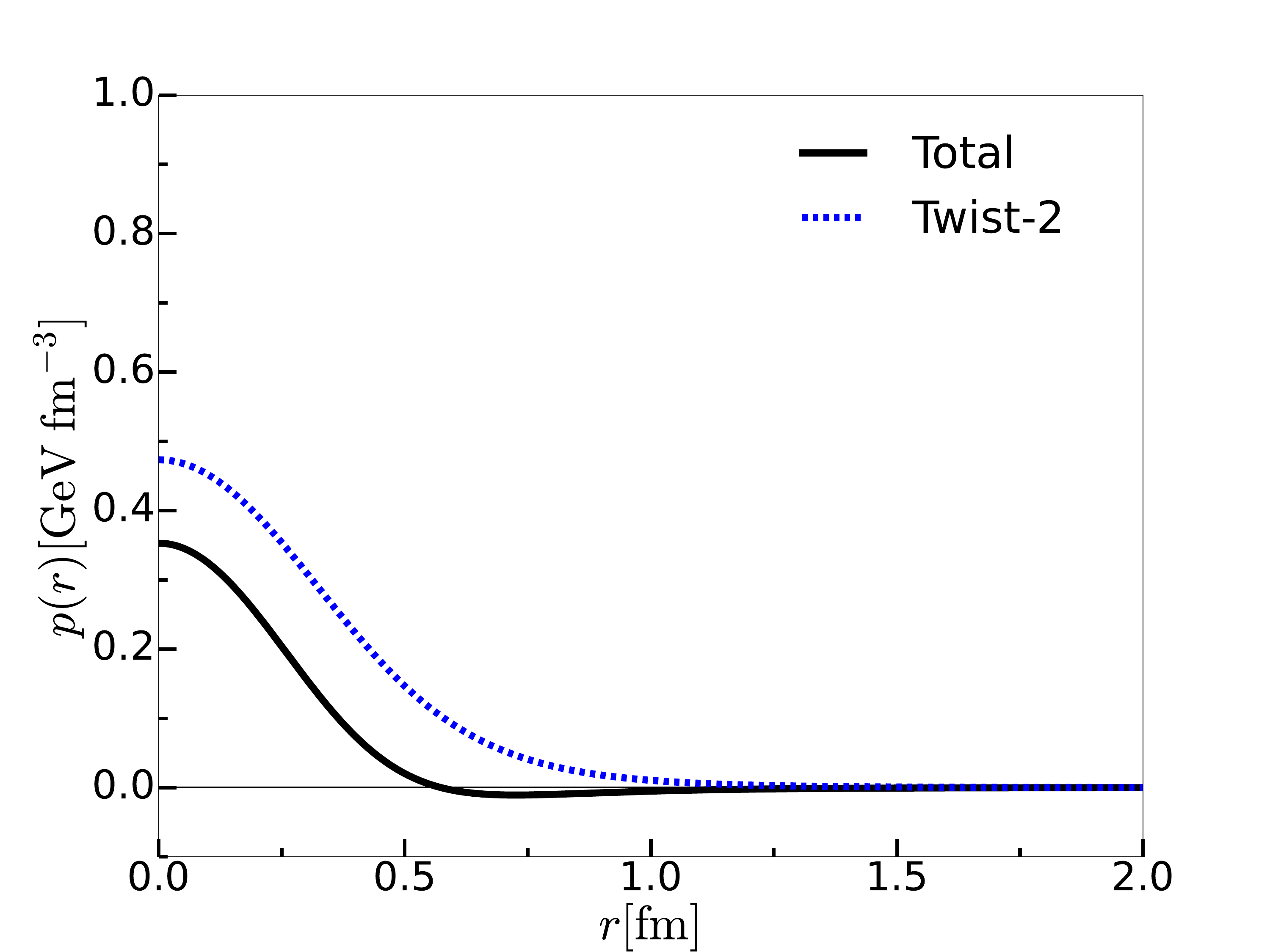}
\includegraphics[width=0.49\textwidth]{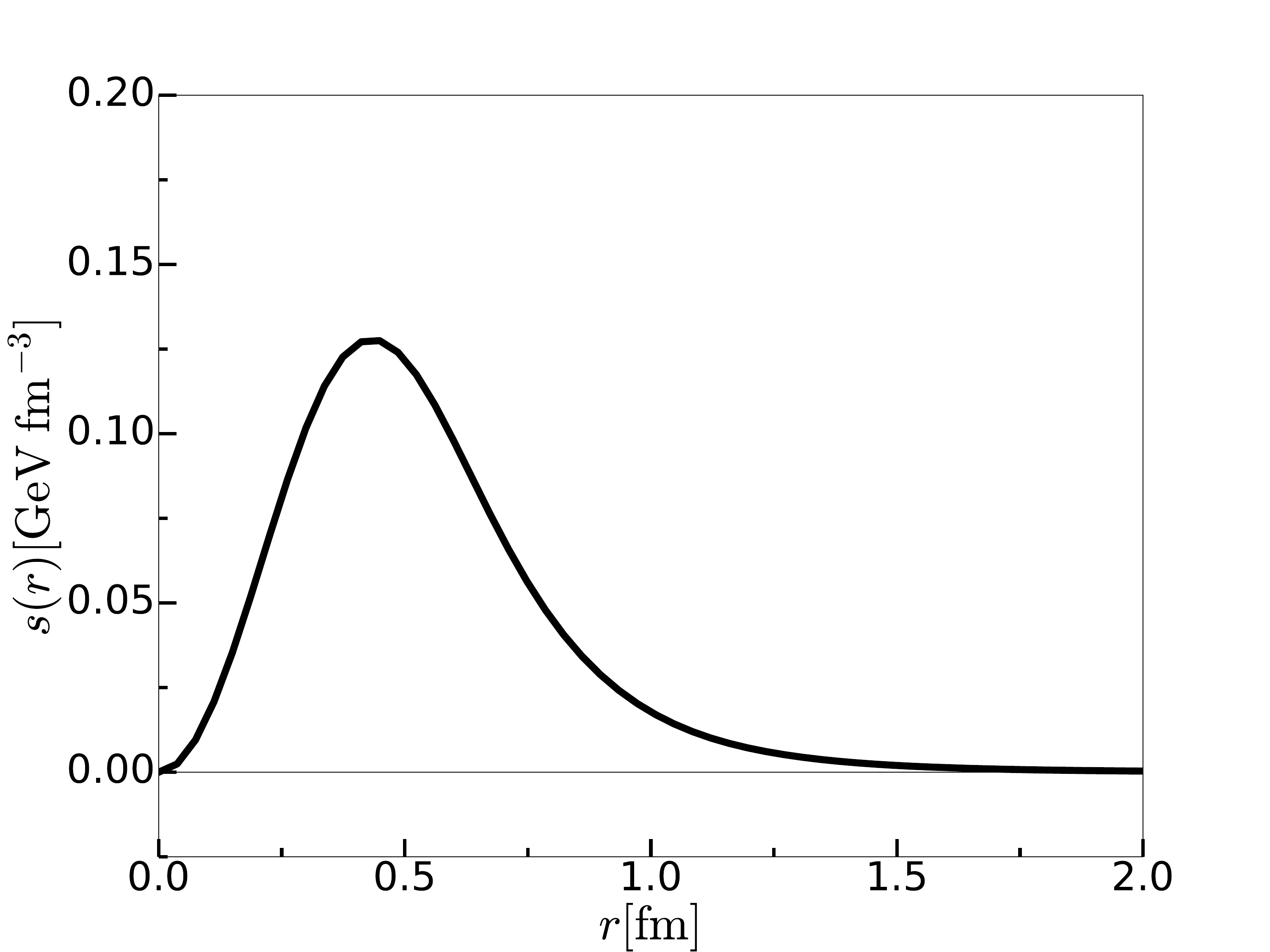}
\caption{The energy, spin, pressure, and shear-force distributions,
  shown from the upper-left to the lower-right panel, respectively.
  Only the energy and pressure distributions can be decomposed into
  the twist-2 and twist-4 contributions.}
\label{fig:3} 
\end{figure}

\begin{figure}[htp] 
\centering 
\includegraphics[width=0.48\textwidth]{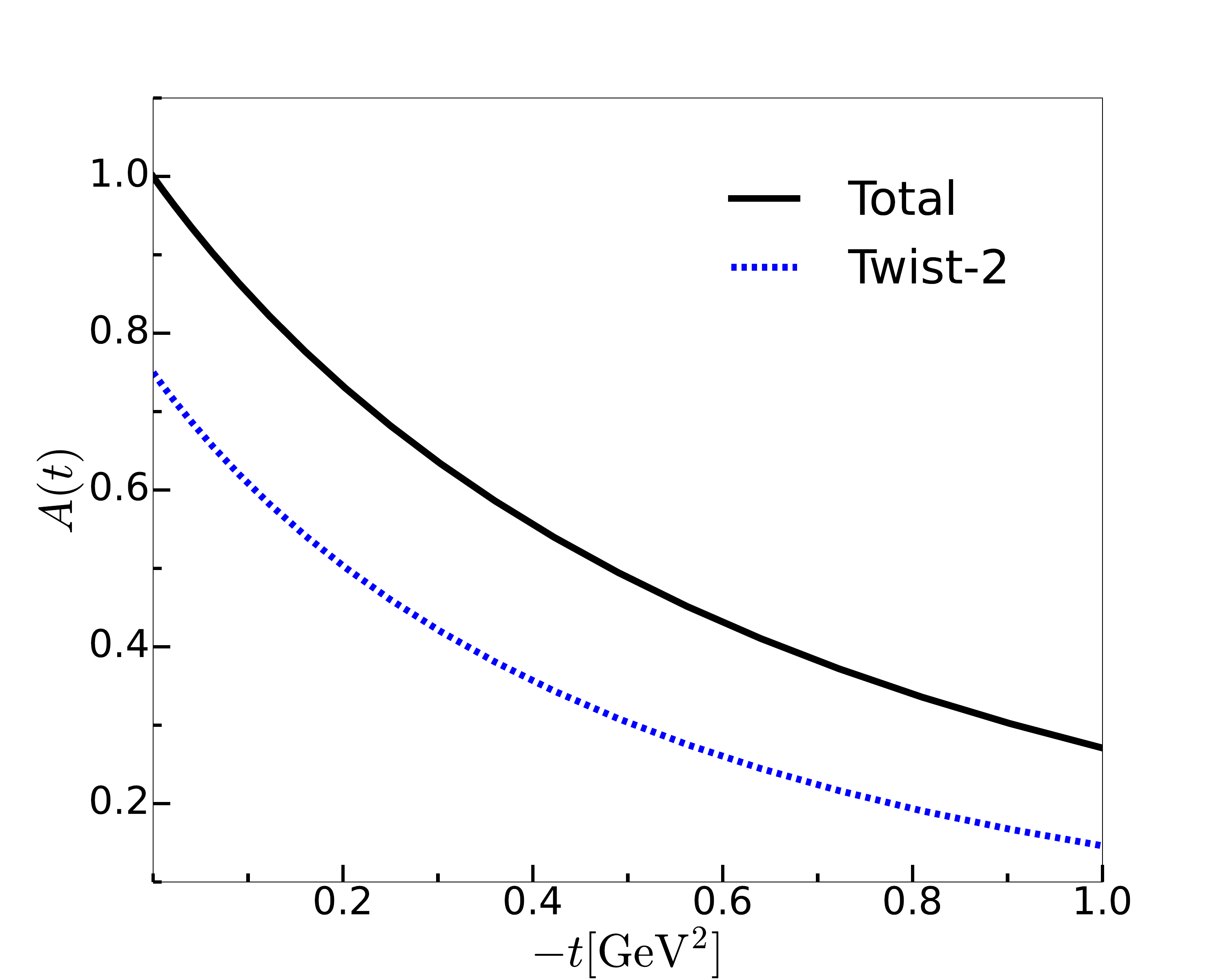}
\includegraphics[width=0.48\textwidth]{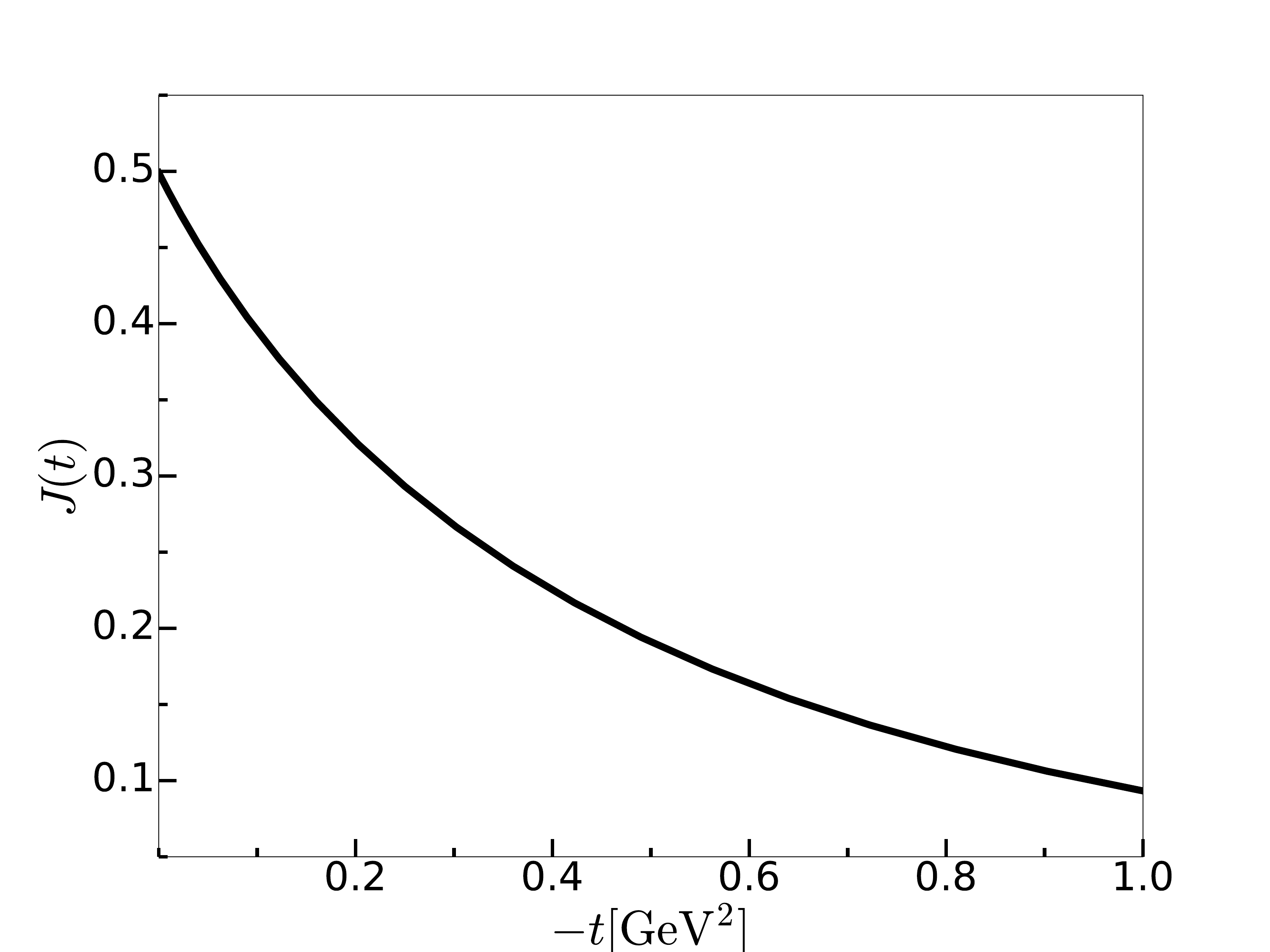}
\includegraphics[width=0.55\textwidth]{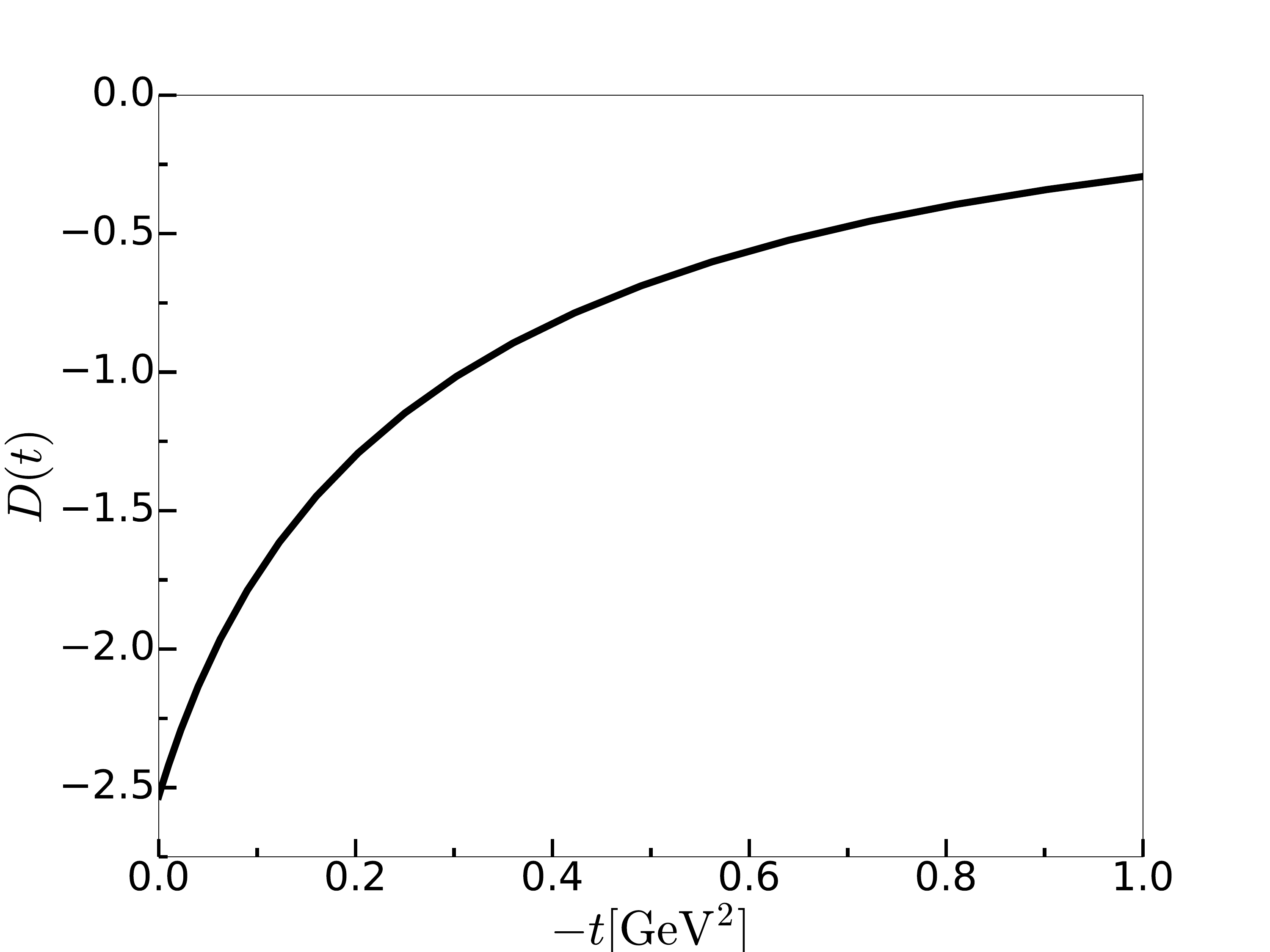}
\caption{The energy, spin, and $D$-term form factors, shown from the
  upper-left to the lower panel, respectively. Only the energy form
  factor can be decomposed into the twist-2 and twist-4 contributions.} 
\label{fig:4} 
\end{figure}
In the present talk, we have focused on the twist-2 contributions to
the GFFs. In Figs.~\ref{fig:3} and \ref{fig:4}, we present the
numerical results for the EMT distributions and the corresponding
GFFs. As discussed above, the twist-4 EMT operator contributes to the
energy and pressure distributions. As a result, only the $A$ form
factor can be decomposed into the twist-2 and twist-4 parts, 
where the total $\bar{c}$ form factor vanishes.

\section{Summary and conclusions}
\label{sec:summary} 

In the present talk, we have reviewed the flavor decomposition of the
gravitational form factors of the proton and the associated mechanical
properties within the chiral quark-soliton model. We first constructed
the effective EMT operator from the QCD instanton vacuum, in which the
gluon degrees of freedom are integrated out and the gluonic
contribution is negligibly small at the leading twist. 

We then presented the numerical results for the flavor-decomposed EMT
distributions and the corresponding GFFs. The twist-2 energy
distribution carries three fourths of the nucleon mass, and the up
quark dominates both the mass and the spin of the proton, whereas the
strange-quark contribution to them is tiny. The total angular momentum
is normalized exactly to $1/2$, as guaranteed by the Ji sum rule. As
for the mechanical properties, the twist-2 pressure is repulsive over
the entire space, and the global stability condition is fulfilled only
when the twist-4 contribution is included. The shear-force
distribution satisfies the positivity condition, which ensures the
local stability of the proton. Notably, the strange quark contributes
sizably to the $D$-term form factor, reflecting the importance of the
sea-quark contributions to the quadrupole structure of the proton.
\section*{Declarations}

\subsection*{Ethics approval and consent to participate}
Not applicable.

\subsection*{Consent for publication}
Not applicable.

\subsection*{Availability of data and materials}
The datasets generated and analysed during the current study are available
from the corresponding author upon reasonable request.

\subsection*{Competing interests}
The authors declare no competing interests.

\subsection*{Funding}
This work was supported by the National Research Foundation of Korea
(NRF) grant funded by the Korea government (MSIT) under Grant
No.~RS-2025-00513982. 

\subsection*{Authors' contributions}
H.-Ch.K. suggested the related ideas and wrote the main manuscript text.
J.Y.K. carried out the theoretical calculation and derived the main results.
H.Y.W. performed all the numerical calculations and prepared all the
figures. All authors reviewed the manuscript.

\subsection*{Acknowledgements}
H.-Ch.K. is grateful to the organizers of APFB 2025 for the invitation to
deliver a plenary talk. 
\backmatter

\bibstyle{sn-mathphys-num} 
\bibliography{apfb}%

@article{Accardi:2012qut,
    author = "Accardi, A. and others",
    editor = "Deshpande, A. and Meziani, Z. E. and Qiu, J. W.",
    title = "{Electron Ion Collider: The Next QCD Frontier}: {Understanding the glue that binds us all}",
    eprint = "1212.1701",
    archivePrefix = "arXiv",
    primaryClass = "nucl-ex",
    reportNumber = "BNL-98815-2012-JA, JLAB-PHY-12-1652",
    doi = "10.1140/epja/i2016-16268-9",
    journal = "Eur. Phys. J. A",
    volume = "52",
    number = "9",
    pages = "268",
    year = "2016"
}

@article{AbdulKhalek:2021gbh,
    author = "Abdul Khalek, R. and others",
    title = "{Science Requirements and Detector Concepts for the Electron-Ion Collider}: {EIC Yellow Report}",
    eprint = "2103.05419",
    archivePrefix = "arXiv",
    primaryClass = "physics.ins-det",
    reportNumber = "BNL-220990-2021-FORE, JLAB-PHY-21-3198, LA-UR-21-20953",
    doi = "10.1016/j.nuclphysa.2022.122447",
    journal = "Nucl. Phys. A",
    volume = "1026",
    pages = "122447",
    year = "2022"
}

@article{Won:2023cyd,
    author = "Won, Ho-Yeon and Kim, Hyun-Chul and Kim, June-Young",
    title = "{Flavor structure of the energy-momentum tensor form factors of the proton}",
    eprint = "2302.02974",
    archivePrefix = "arXiv",
    primaryClass = "hep-ph",
    reportNumber = "INHA-NTG-02/2023, JLAB-THY-23-3754",
    doi = "10.1016/j.physletb.2024.138489",
    journal = "Phys. Lett. B",
    volume = "850",
    pages = "138489",
    year = "2024"
}

@article{Balla:1997hf,
    author = "Balla, J. and Polyakov, Maxim V. and Weiss, C.",
    title = "{Nucleon matrix elements of higher twist operators from the instanton vacuum}",
    eprint = "hep-ph/9707515",
    archivePrefix = "arXiv",
    reportNumber = "RUB-TPII-6-97",
    doi = "10.1016/S0550-3213(98)00638-5",
    journal = "Nucl. Phys. B",
    volume = "510",
    pages = "327--364",
    year = "1998"
}

@article{Won:2023ial,
    author = "Won, Ho-Yeon and Kim, Hyun-Chul and Kim, June-Young",
    title = "{Role of strange quarks in the $D$-term and cosmological constant term of the proton}",
    eprint = "2307.00740",
    archivePrefix = "arXiv",
    primaryClass = "hep-ph",
    reportNumber = "INHA-NTG-04/2023",
    month = "7",
    year = "2023"
}

@article{Wakamatsu:2006dy,
    author = "Wakamatsu, M. and Nakakoji, Y.",
    title = "{Generalized form factors, generalized parton distributions and the spin contents of the nucleon}",
    eprint = "hep-ph/0605279",
    archivePrefix = "arXiv",
    reportNumber = "OU-HET-561",
    doi = "10.1103/PhysRevD.74.054006",
    journal = "Phys. Rev. D",
    volume = "74",
    pages = "054006",
    year = "2006"
}

@article{Wakamatsu:2005vk,
    author = "Wakamatsu, M. and Tsujimoto, H.",
    title = "{The Generalized parton distribution functions and the nucleon spin sum rules in the chiral quark soliton model}",
    eprint = "hep-ph/0502030",
    archivePrefix = "arXiv",
    reportNumber = "OU-HEP-518",
    doi = "10.1103/PhysRevD.71.074001",
    journal = "Phys. Rev. D",
    volume = "71",
    pages = "074001",
    year = "2005"
}

@article{Muller:1994ses,
    author = {M\"uller, Dieter and Robaschik, D. and Geyer, B. and Dittes, F. -M. and Ho\v{r}ej\v{s}i, J.},
    title = "{Wave functions, evolution equations and evolution kernels from light ray operators of QCD}",
    eprint = "hep-ph/9812448",
    archivePrefix = "arXiv",
    reportNumber = "NTZ-6-91, NTZ-91-6",
    doi = "10.1002/prop.2190420202",
    journal = "Fortsch. Phys.",
    volume = "42",
    pages = "101--141",
    year = "1994"
}

@article{Radyushkin:1996nd,
    author = "Radyushkin, A. V.",
    title = "{Scaling limit of deeply virtual Compton scattering}",
    eprint = "hep-ph/9604317",
    archivePrefix = "arXiv",
    reportNumber = "CEBAF-TH-96-05",
    doi = "10.1016/0370-2693(96)00528-X",
    journal = "Phys. Lett. B",
    volume = "380",
    pages = "417--425",
    year = "1996"
}

@article{Ji:1996ek,
    author = "Ji, Xiang-Dong",
    title = "{Gauge-Invariant Decomposition of Nucleon Spin}",
    eprint = "hep-ph/9603249",
    archivePrefix = "arXiv",
    reportNumber = "MIT-CTP-2517",
    doi = "10.1103/PhysRevLett.78.610",
    journal = "Phys. Rev. Lett.",
    volume = "78",
    pages = "610--613",
    year = "1997"
}

@article{Kumericki:2019ddg,
    author = "Kumeri\v{c}ki, Kre\v{s}imir",
    title = "{Measurability of pressure inside the proton}",
    doi = "10.1038/s41586-019-1211-6",
    journal = "Nature",
    volume = "570",
    number = "7759",
    pages = "E1--E2",
    year = "2019"
}

@article{Burkert:2018bqq,
    author = "Burkert, V. D. and Elouadrhiri, L. and Girod, F. X.",
    title = "{The pressure distribution inside the proton}",
    doi = "10.1038/s41586-018-0060-z",
    journal = "Nature",
    volume = "557",
    number = "7705",
    pages = "396--399",
    year = "2018"
}

@article{Burkert:2021ith,
    author = "Burkert, V. D. and Elouadrhiri, L. and Girod, F. X.",
    title = "{Determination of shear forces inside the proton}",
    eprint = "2104.02031",
    archivePrefix = "arXiv",
    journal = "",    
    primaryClass = "nucl-ex",
    month = "4",
    year = "2021"
}

@article{Diakonov:1996sr,
    author = "Diakonov, Dmitri and Petrov, V. and Pobylitsa, P. and Polyakov, Maxim V. and Weiss, C.",
    title = "{Nucleon parton distributions at low normalization point in the large N(c) limit}",
    eprint = "hep-ph/9606314",
    archivePrefix = "arXiv",
    reportNumber = "RUB-TPII-9-96",
    doi = "10.1016/S0550-3213(96)00486-5",
    journal = "Nucl. Phys. B",
    volume = "480",
    pages = "341--380",
    year = "1996"
}

@article{Diakonov:1997vc,
    author = "Diakonov, Dmitri and Petrov, V. Yu. and Pobylitsa, P. V. and Polyakov, Maxim V. and Weiss, C.",
    title = "{Unpolarized and polarized quark distributions in the large N(c) limit}",
    eprint = "hep-ph/9703420",
    archivePrefix = "arXiv",
    reportNumber = "NORDITA-97-24-P, RUB-TPII-3-97",
    doi = "10.1103/PhysRevD.56.4069",
    journal = "Phys. Rev. D",
    volume = "56",
    pages = "4069--4083",
    year = "1997"
}

@article{Perevalova:2016dln,
    author = "Perevalova, I. A. and Polyakov, M. V. and Schweitzer, P.",
    title = "{On LHCb pentaquarks as a baryon-$\psi$(2S) bound state: prediction of isospin-$\frac3{2}$ pentaquarks with hidden charm}",
    eprint = "1607.07008",
    archivePrefix = "arXiv",
    primaryClass = "hep-ph",
    doi = "10.1103/PhysRevD.94.054024",
    journal = "Phys. Rev. D",
    volume = "94",
    number = "5",
    pages = "054024",
    year = "2016"
}

@article{Won:2023zmf,
    author = "Won, Ho-Yeon and Kim, Hyun-Chul and Kim, June-Young",
    title = "{Mechanical structure of the nucleon and the baryon octet: twist-2 case}",
    eprint = "2310.04670",
    archivePrefix = "arXiv",
    primaryClass = "hep-ph",
    reportNumber = "JLAB-THY-23-3938",
    doi = "10.1007/JHEP05(2024)173",
    journal = "JHEP",
    volume = "05",
    pages = "173",
    year = "2024"
}

@article{Polyakov:2018exb,
    author = "Polyakov, Maxim V. and Son, Hyeon-Dong",
    title = "{Nucleon gravitational form factors from instantons: forces between quark and gluon subsystems}",
    eprint = "1808.00155",
    archivePrefix = "arXiv",
    primaryClass = "hep-ph",
    doi = "10.1007/JHEP09(2018)156",
    journal = "JHEP",
    volume = "09",
    pages = "156",
    year = "2018"
}

@article{Polyakov:2002yz,
    author = "Polyakov, M. V.",
    title = "{Generalized parton distributions and strong forces inside nucleons and nuclei}",
    eprint = "hep-ph/0210165",
    archivePrefix = "arXiv",
    reportNumber = "RUB-TP2-14-02",
    doi = "10.1016/S0370-2693(03)00036-4",
    journal = "Phys. Lett. B",
    volume = "555",
    pages = "57--62",
    year = "2003"
}

@article{Polyakov:2018zvc,
    author = "Polyakov, Maxim V. and Schweitzer, Peter",
    title = "{Forces inside hadrons: pressure, surface tension, mechanical radius, and all that}",
    eprint = "1805.06596",
    archivePrefix = "arXiv",
    primaryClass = "hep-ph",
    doi = "10.1142/S0217751X18300259",
    journal = "Int. J. Mod. Phys. A",
    volume = "33",
    number = "26",
    pages = "1830025",
    year = "2018"
}

@article{Pagels:1966zza,
    author = "Pagels, Heinz",
    title = "{Energy-Momentum Structure Form Factors of Particles}",
    doi = "10.1103/PhysRev.144.1250",
    journal = "Phys. Rev.",
    volume = "144",
    pages = "1250--1260",
    year = "1966"
}

@article{Kobzarev:1962wt,
    author = "Kobzarev, I. Yu. and Okun, L. B.",
    title = "{GRAVITATIONAL INTERACTION OF FERMIONS}",
    journal = "Zh. Eksp. Teor. Fiz.",
    volume = "43",
    pages = "1904--1909",
    year = "1962"
}

@article{Wakamatsu:2007uc,
    author = "Wakamatsu, M.",
    title = "{On the D-term of the nucleon generalized parton distributions}",
    eprint = "hep-ph/0701057",
    archivePrefix = "arXiv",
    reportNumber = "OU-HET-572",
    doi = "10.1016/j.physletb.2007.03.013",
    journal = "Phys. Lett. B",
    volume = "648",
    pages = "181--185",
    year = "2007"
}

@article{Kim:2019gka,
    author = "Kim, June-Young and Kim, Hyun-Chul",
    title = "{Electromagnetic form factors of the baryon decuplet with flavor SU(3) symmetry breaking}",
    eprint = "1905.04017",
    archivePrefix = "arXiv",
    primaryClass = "hep-ph",
    reportNumber = "INHA-NTG-06/2019",
    doi = "10.1140/epjc/s10052-019-7079-7",
    journal = "Eur. Phys. J. C",
    volume = "79",
    number = "7",
    pages = "570",
    year = "2019"
}
 
\end{document}